\documentclass[showpacs,10pt,twocolumn,prb]{revtex4-1}

\usepackage{amsmath}
\usepackage{amssymb}
\usepackage{graphicx}
\usepackage{amssymb}
\usepackage{graphics}
\usepackage{epsfig}
\usepackage{CJK}
\usepackage{color}

\begin{document}

\begin{CJK*}{GBK}{Song}
\title{Three-dimensional magnetic critical behavior in CrI$_3$}
\author{Yu Liu and C. Petrovic}
\affiliation{Condensed Matter Physics and Materials Science Department, Brookhaven National Laboratory, Upton, New York 11973, USA}
\date{\today}

\begin{abstract}
CrI$_3$ is a promising candidate for the van der Waals bonded ferromagnetic devices since its ferromagnetism can be maintained upon exfoliating of bulk crystals down to single layer. In this work we studied critical properties of bulk CrI$_3$ single crystals around the paramagnetic to ferromagnetic phase transition. Critical exponents $\beta$ = 0.260(4) with a critical temperature $T_c$ = 60.05(13) K and $\gamma$ = 1.136(6) with $T_c$ = 60.43(4) K are obtained by the Kouvel-Fisher method, whereas $\delta$ = 5.32(2) is obtained by a critical isotherm analysis at $T_c$ = 60 K. The critical exponents determined in bulk CrI$_3$ single crystals suggest a three-dimensional long-range magnetic coupling with the exchange distance decaying as $J(r)\approx r^{-4.69}$.
\end{abstract}

\maketitle
\end{CJK*}

\section{INTRODUCTION}

Ferromagnetic (FM) semiconductors, possessing both ferromagnetism and semiconducting character, form the basis for spintronics application. In order to develop the next-generation nano-spintronic devices, low dimensional intrinsically FM semiconductors are needed.\cite{McGuire0, McGuire, Huang, Gong}

Recently, Cr$_2$X$_2$Te$_6$ (X = Si, Ge, Sn) have been identified as promising candidates for long-range magnetism in nanosheets.\cite{Gong, Lin, Zhuang} Bulk Cr$_2$X$_2$Te$_6$ are small band gap FM semiconductors with the Curie temperature ($T_c$) of 32 K for Cr$_2$Si$_2$Te$_6$ and 61 K for Cr$_2$Ge$_2$Te$_6$, respectively.\cite{Ouvrard, Carteaux1, Carteaux2, Casto, Zhang} The theoretical calculation based on a Heisenberg model predicts robust 2D ferromagnetism in monolayer with $T_c$ $\sim$ 35.7 K for Cr$_2$Si$_2$Te$_6$ and $\sim$ 57.2 K for Cr$_2$Ge$_2$Te$_6$,\cite{Li} when the nearest-neighbor (NN) exchange is only considered. When the second and the third NN exchange interactions are also taken into consideration, the monolayer Cr$_2$Si$_2$Te$_6$ is expected to be antiferromagnetic (AFM), whereas Cr$_2$Ge$_2$Te$_6$ is still FM with $T_c$ of 106 K.\cite{Sivadas} However, the scanning magneto-optic Kerr microscopy experiment shows that the $T_c$ monotonically decreases with decreasing thickness of Cr$_2$Ge$_2$Te$_6$, from bulk of 68 K to a bilayer value of 30 K, and the FM order is not present in a single layer of Cr$_2$Ge$_2$Te$_6$,\cite{Gong} different from the theoretical prediction.

In distinct contrast to Cr$_2$Ge$_2$Te$_6$, CrI$_3$ displays a similar $T_c$ of 61 K and significant magnetic anisotropy but the FM order is also present in single layer with $T_c$ of 45 K.\cite{Huang} The monolayer CrI$_3$ could be well described by the Ising model.\cite{Huang, Griffiths} The rich magnetic phase diagram, including in-plane AFM, off-plane FM, and in-plane FM phases, is also predicted by applying lateral strain and/or charge doping.\cite{Zheng} Additionally, Huang et al. demonstrated that the magnetism in CrI$_3$ is strongly layer-dependent, from FM in the monolayer, to AFM in the bilayer, and back to FM in the trilayer,\cite{Huang} providing great opportunities for designing new magneto-optoelectronic devices.

In the present work we focus on the nature of the FM transition in bulk CrI$_3$ single crystals, which has not been studied. The critical behavior has been investigated by modified Arrott plot, Kouvel-Fisher plot, and critical isotherm analysis. The determined critical exponents $\beta$ = 0.260(4) with $T_c$ = 60.05(13) K, $\gamma$ = 1.136(6) with $T_c$ = 60.43(4) K, and $\delta$ = 5.32(2) at $T_c$ = 60 K do not belong to any single universality class but are in what between is expected for three-dimensional (3D) Ising model ($\beta = 0.325, \gamma = 1.24$, and $\delta$ = 4.82) and tricritical mean-field model ($\beta = 0.25, \gamma = 1.0$, and $\delta$ = 5). The magnetic exchange distance is found to decay as $J(r)\approx r^{-4.69}$.

\section{EXPERIMENTAL DETAILS}

Bulk CrI$_3$ single crystals were grown by chemical vapor transport (CVT) starting from an intimate mixture of pure elements chromium powder (99.95 $\%$, Alfa Aesar) and anhydrous iodine beads (99.99 $\%$, Alfa Aesar) with a molar ratio of 1 : 3. The starting materials were sealed in an evacuated quartz tube and then placed inside a multi-zone furnace reacted over a period of 7 days with source zone at 650 $^\circ$C, middle growth zone at 550 $^\circ$C, and third zone at 600 $^\circ$C. The x-ray diffraction (XRD) data were taken with Cu K$_{\alpha}$ ($\lambda=0.15418$ nm) radiation of Rigaku Miniflex powder diffractometer. The element analysis was performed using an energy-dispersive x-ray spectroscopy in a JEOL LSM-6500 scanning electron microscope (SEM), confirming a stoichiometric CrI$_3$ single crystal. The magnetization was measured in Quantum Design MPMS-XL5 system. The isothermal $M(H)$ curves were measured in $\Delta T$ = 1 K intervals. The applied field ($H_a$) has been corrected for the internal field as $H = H_a - NM$, where $M$ is the measured magnetization and $N$ is the demagnetization factor. The corrected $H$ was used for the analysis of critical behavior.

\section{RESULTS AND DISCUSSIONS}

\begin{figure}
\centerline{\includegraphics[scale=1]{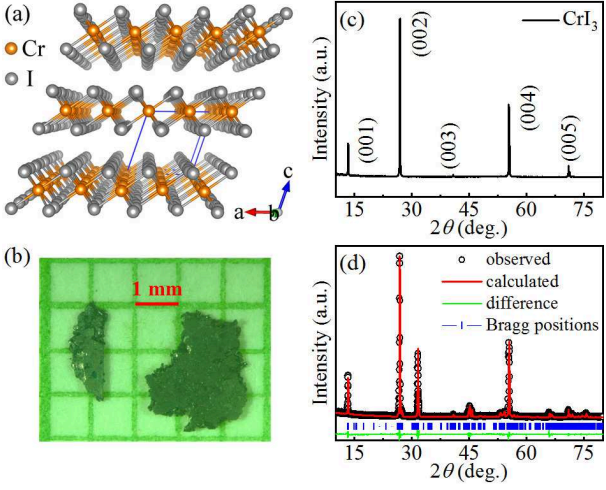}}
\caption{(Color online). (a) Crystal structure of CrI$_3$ at room temperature. (b) Image of representative single crystals. (c) Single-crystal x-ray diffraction (XRD) and (d) powder XRD patterns of CrI$_3$ at room temperature. The vertical tick marks represent Bragg reflections of the $C2/m$ space group.}
\label{XRD}
\end{figure}

Figure 1(a) shows the crystal structure of CrI$_3$ at room temperature formed in the monoclinic AlCl$_3$ type.\cite{McGuire} The Cr ions are arranged in a honeycomb network and located at the centers of edge-sharing octahedra of six I atoms, which are each bonded to two Cr ions. The sandwich-like I-Cr-I triple layers of composition CrI$_3$ are stacked along the $c$ axis with van der Waals (vdW) gaps separating them. The as-grown single crystals are formed as thin and flexible platelets with irregular shapes, as shown in Fig. 1(b). In the single-crystal XRD scan [Fig. 1(c)], only $(00l)$ peaks are detected, indicating the crystal surface is normal to the $c$ axis with the plate-shaped surface parallel to the $ab$ plane. Figure 1(d) shows the powder XRD pattern of CrI$_3$, in which the observed peaks are well fitted with the $C2/m$ space group. The determined lattice parameters are $a = 6.866(2)$ {\AA}, $b = 11.856(2)$ {\AA}, $c = 6.996(2)$ {\AA}, and $\beta = 108.68^\circ$,  which are very close to the values reported previously.\cite{McGuire}

\begin{figure}
\centerline{\includegraphics[scale=0.9]{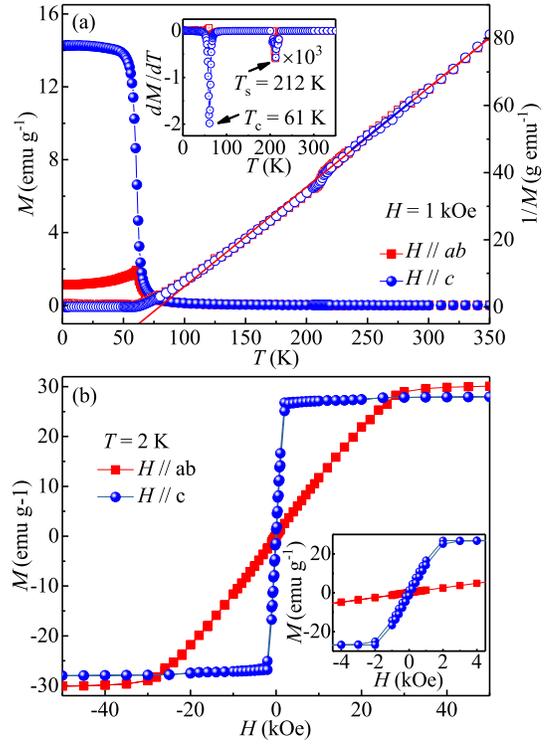}}
\caption{(Color online). (a) Temperature dependence of magnetization for CrI$_3$ measured in the magnetic field $H$ = 1 kOe applied in the $ab$ plane and along the $c$ axis. The solid lines are fitted by the modified Curie-Weiss law $\chi = \chi_0 + \frac{C}{T-\theta}$, where $\chi_0$ is the temperature-independent susceptibility, $C$ is the Curie-Weiss constant, and $\theta$ is the Weiss temperature. Inset: $dM/dT$ vs $T$. (b) Field dependence of magnetization for CrI$_3$ measured at $T$ = 2 K. Inset: the magnification of the low field region.}
\label{MTH}
\end{figure}

\begin{figure}
\centerline{\includegraphics[scale=1]{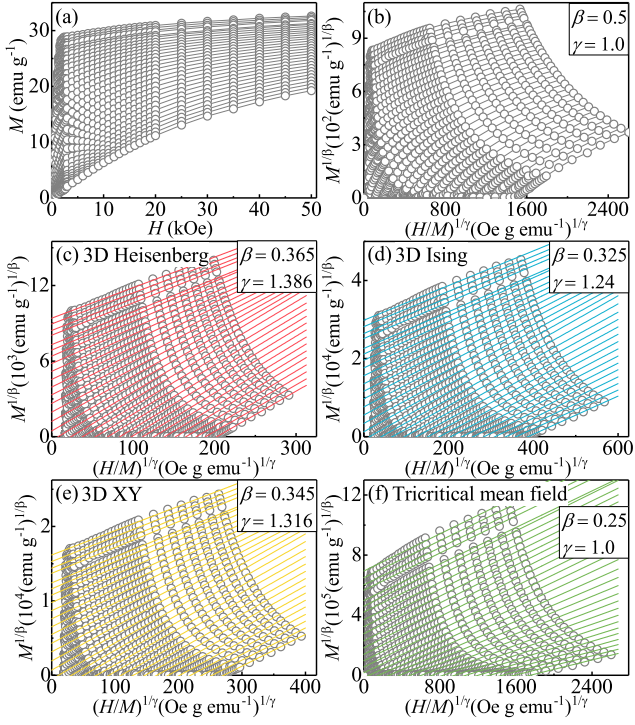}}
\caption{(Color online). (a) Typical initial isothermal magnetization curves measured along the $c$ axis from $T$ = 45 K to $T$ = 75 K for CrI$_3$. (b) Arrott plots of $M^2$ vs $H/M$. The $M^{1/\beta}$ vs $(H/M)^{1/\gamma}$ with parameters of (c) 3D Heisenberg model, (d) 3D Ising model, (e) 3D XY model, and (f) Tricritical mean-field model. The straight lines are the linear fit of isothermals at different temperatures.}
\label{Arrot}
\end{figure}

\begin{figure}
\centerline{\includegraphics[scale=0.9]{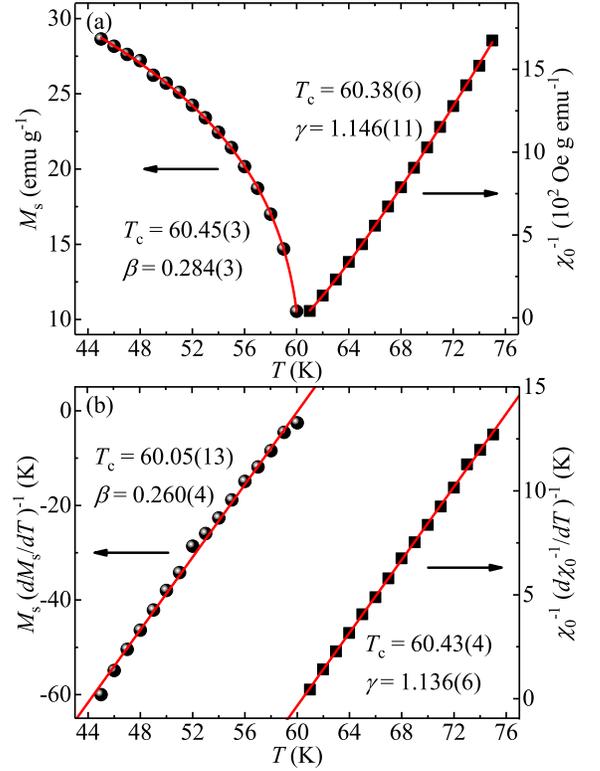}}
\caption{(Color online). (a) Temperature dependence of the spontaneous magnetization $M_s$ (left) and the inverse initial susceptibility $\chi_0^{-1}$ (right) with solid fitting curves for CrI$_3$. (b) Kouvel-Fisher plots of $M_s(dM_s/dT)^{-1}$ (left) and $\chi_0^{-1}(d\chi_0^{-1}/dT)^{-1}$ (right) with solid fitting curves for CrI$_3$.}
\label{KF}
\end{figure}

A clear paramagnetic (PM) to FM transition is observed in magnetic susceptibility [Fig. 2(a)]. The crystallographic $c$ axis is found to be the easy axis. Critical temperature $T_c \approx 61$ K is roughly determined from the minimum of the $dM/dT$ curve [Fig. 2(a) inset], in agreement with previous work.\cite{McGuire} Additionally, another magnetization anomaly observed at $T_s \approx 212$ K is corresponding to the crystallographic phase transition from rhombohedral ($R\bar{3}$) structure to monoclinic ($C2/m$) structure upon warming, which was well studied in previous report.\cite{McGuire} Linear fit of the $1/M$ data in the temperature range of 225-350 K [Fig. 2(a)] yields the Weiss temperature $\theta_{ab} \approx 76.4(8)$ K or $\theta_c \approx 79.4(17)$ K, i.e. dominant FM exchange interactions. The effective moment $\mu_{\textrm{eff}}$ = 3.44(1) $\mu_B$/Cr obtained from $H//ab$ data is identical to $\mu_{\textrm{eff}}$ = 3.42(1) $\mu_B$/Cr from $H//c$ data, indicating a nearly isotropic PM behavior at high temperatures. The value of $\mu_{\textrm{eff}}$ is also close to the theoretical value expected for Cr$^{3+}$ of 3.87 $\mu_B$. Isothermal magnetization measured at $T$ = 2 K [Fig. 2(b)] shows saturation field $H_s \approx 2 $ kOe for $H//c$, much smaller than $H_s \approx 30 $ kOe for $H//ab$, thus confirming the easy $c$ axis. The saturation moment at $T$ = 2 K is $M_s \approx$ 2.21(2) $\mu_B$/Cr for $H//ab$ and $M_s \approx$ 2.14(1) $\mu_B$/Cr for $H//c$, respectively. The $M(H)$ in low field region [Fig. 2(b) inset] reveals small hysteresis with the coercive forces $H_{ab} = 100$ Oe for $H // ab$ and $H_c = 85$ Oe for $H // c$, respectively, in agreement previous report.\cite{McGuire}

A second-order phase transition criticality is characterized with interdependent critical exponents.\cite{Stanley} Near second-order phase transition the correlation length diverges as $\xi = \xi_0 |(T-T_c)/T_c|^{-\nu}$ and there are universal scaling laws for the spontaneous magnetization $M_s$ and the inverse initial magnetic susceptibility $\chi_0^{-1}$. The $M_s$ below $T_c$, the $\chi_0^{-1}$ above $T_c$, and the $M(H)$ at $T_c$ are characterized by critical exponents $\beta$, $\gamma$, and $\delta$ that give insight into magnetic interactions, correlating length, spin-dimensionality, and decaying distance of magnetic coupling,
\begin{equation}
M_s (T) = M_0(-\varepsilon)^\beta, \varepsilon < 0, T < T_c,
\end{equation}
\begin{equation}
\chi_0^{-1} (T) = (h_0/m_0)\varepsilon^\gamma, \varepsilon > 0, T > T_c,
\end{equation}
\begin{equation}
M = DH^{1/\delta}, \varepsilon = 0, T = T_c,
\end{equation}
where $\varepsilon = (T-T_c)/T_c$ is the reduced temperature, and $M_0$, $h_0/m_0$ and $D$ are the critical amplitudes.\cite{Fisher} The magnetic equation of state can be expressed as
\begin{equation}
M(H,\varepsilon) = \varepsilon^\beta f_\pm(H/\varepsilon^{\beta+\gamma}),
\end{equation}
where $f_+$ for $T>T_c$ and $f_-$ for $T<T_c$, respectively, are the regular functions. Eq.(4) can be written in terms of renormalized magnetization $m\equiv\varepsilon^{-\beta}M(H,\varepsilon)$ and renormalized field $h\equiv\varepsilon^{-(\beta+\gamma)}H$ as
\begin{equation}
m = f_\pm(h).
\end{equation}
This suggests that for true scaling relations and the right choice of $\beta$, $\gamma$, and $\delta$ values, scaled $m$ and $h$ will fall on universal curves above $T_c$ and below $T_c$, respectively.

Isothermal magnetization in the temperature range from $T$ = 45 K to $T$ = 75 K is shown in Fig. 3(a). The Arrott plot involves meain-field critical exponents $\beta$ = 0.5 and $\gamma$ = 1.0.\cite{Arrott1} Based on this, magnetization isotherms $M^2$ vs $H/M$ are a set of parallel straight lines. The isotherm at the critical temperature $T_c$ should pass through the origin. This plot gives $\chi_0^{-1}(T)$ and $M_s(T)$ as the intercepts on the $H/M$ axis and positive $M^2$ axis, respectively. As shown in Fig. 3(b), all curves in the Arott plot of CrI$_3$ are nonlinear, with a downward curvature. It demonstrates that the Landau mean-field model is not applicable to CrI$_3$. Whatever, it is possible to estimate the order of the magnetic transition through the slope of the straight line based on Banerjee$^\prime$s criterion.\cite{Banerjee} First (second) order phase transition corresponds to negative (positive) slope. Therefore, the downward slope reveals a second-order PM-FM transition in CrI$_3$.

\begin{figure}
\centerline{\includegraphics[scale=0.9]{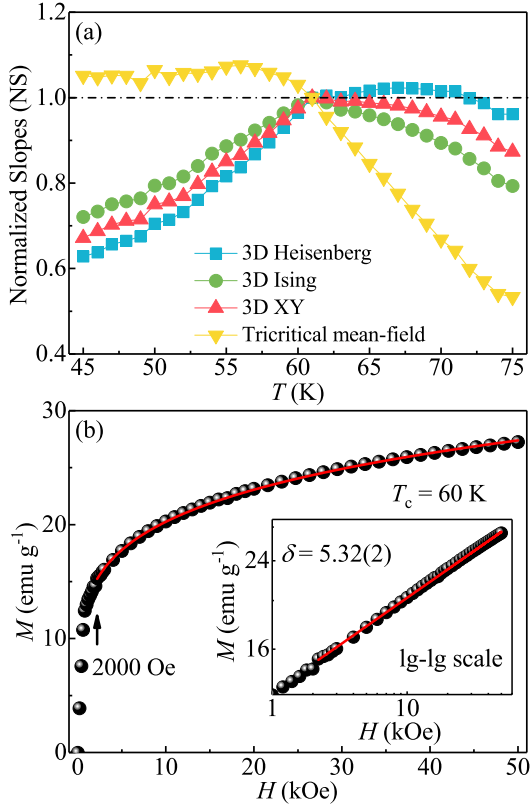}}
\caption{(Color online). (a) Temperature dependence of the normalized slopes $NS = S(T)/S(T_c)$ for different models. (b) Isotherm $M(H)$ collected at $T_c$ = 60 K. Inset: the same plot in log-log scale with a solid fitting curve.}
\label{NS}
\end{figure}

\begin{figure}
\centerline{\includegraphics[scale=0.9]{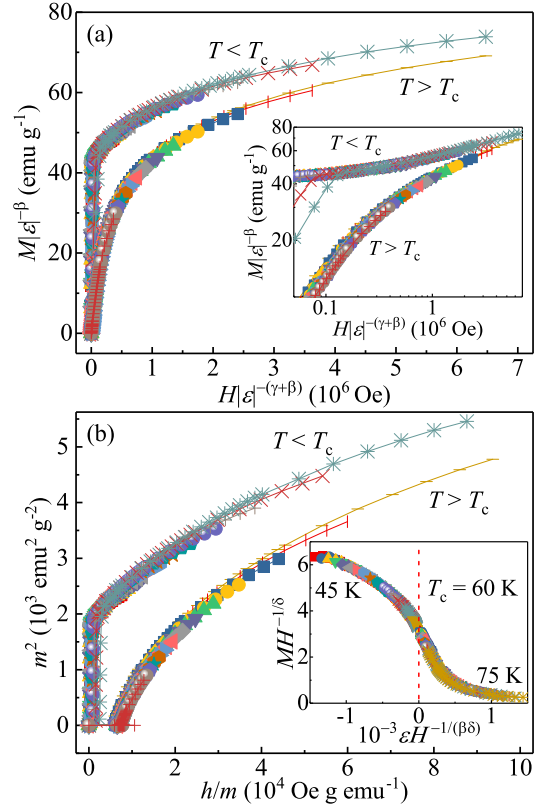}}
\caption{(Color online). (a) Scaling plots of renormalized magnetization $m$ vs renormalized field $h$ below and above $T_c$ for CrI$_3$. Inset: the same plots in log-log scale. (b) Plots of $m^2$ vs $h/m$ for CrI$_3$. Inset: the rescaling of the $M(H)$ curves by $MH^{-1/\delta}$ vs $\varepsilon H^{-1/(\beta\delta)}$.}
\label{renomalized}
\end{figure}

A modified Arrott plot given by the Arrot-Noaks equation of state could give further insight,\cite{Arrott2}
\begin{equation}
(H/M)^{1/\gamma} = a\varepsilon+bM^{1/\beta},
\end{equation}
where $\varepsilon = (T-T_c)/T_c$ is the reduced temperature, and $a$ and $b$ are constants. Four kinds of possible exponents belonging to 3D Heisenberg model ($\beta = 0.365, \gamma = 1.386$), 3D Ising model ($\beta = 0.325, \gamma = 1.24$), 3D XY model ($\beta = 0.345, \gamma = 1.316$),  and tricritical mean-field model ($\beta = 0.25, \gamma = 1.0$) are used to construct the modified Arrott plots,\cite{LeGuillou} as shown in Figs. 3(c)-(f). It is noted that at low field region, the replotted isotherms are slightly curved as they represent averaging over domains magnetized in different directions.\cite{Stanley} Nevertheless, in high field region, all these four constructions exhibit quasi straight lines, suggesting a possible 3D magnetic behavior in bulk CrI$_3$.

We use an iterative method to obtain the exact critical exponents $\beta$ and $\gamma$.\cite{Pramanik} The linear extrapolation from the high field region to the intercepts with the axis $M^{1/\beta}$ and $(H/M)^{1/\gamma}$ yields reliable values of $M_s(T)$ and $\chi_0^{-1}(T)$. A set of $\beta$ and $\gamma$ can be obtained by fitting the data following the Eqs. (1) and (2). New values of $\beta$ and $\gamma$ are then used to reconstruct a modified Arrott plot. Consequently, new $M_s(T)$ and $\chi_0^{-1}(T)$ are generated from the linear extrapolation from the high field region. Therefore, another set of $\beta$ and $\gamma$ can be generated. This procedure is repeated until $\beta$ and $\gamma$ are stable. Then the obtained critical exponents from this method are independent on the initial parameters, confirming reliability. Figure 4(a) presents the final $M_s(T)$ and $\chi_0^{-1}(T)$ with solid fitting curves. The critical exponents $\beta = 0.284(3)$, with $T_c = 60.45(3)$ K, and $\gamma = 1.146(11)$, with $T_c = 60.38(6)$ K, are obtained.

\begin{table*}
\caption{\label{tab}Comparison of critical exponents of CrI$_3$ and Cr$_2$(Si/Ge)$_2$Te$_6$ with different theoretical models.}
\begin{ruledtabular}
\begin{tabular}{lllllll}
  Composition & Theoretical model & Reference & Technique & $\beta$ & $\gamma$ & $\delta$ \\
  \hline
  CrI$_3$ && This work & Modified Arrott plot & 0.284(3) & 1.146(11) & 5.04(1) \\
  && This work & Kouvel-Fisher plot & 0.260(4) & 1.136(6) & 5.37(4) \\
  && This work &Critical isotherm  &   &   & 5.32(2) \\
  Cr$_2$Si$_2$Te$_6$ && \cite{BJLiu} & Kouvel-Fisher plot & 0.175(9) & 1.562(9) & 9.925(56) \\
  Cr$_2$Ge$_2$Te$_6$ && \cite{YuLiu} & Kouvel-Fisher plot & 0.200(3) & 1.28(3) & 7.40(5) \\
  & 3D Heisenberg & \cite{Kaul} & Theory & 0.365 & 1.386 & 4.8 \\
  Cu$_2$OSeO$_3$ && \cite{Zivkovic} & AC susceptibility & 0.37(1) & 1.44(4) & 4.9(1)\\
  & 3D XY & \cite{Kaul} & Theory & 0.345 & 1.316 & 4.81 \\
  & 3D Ising & \cite{Kaul} & Theory & 0.325 & 1.24 & 4.82 \\
  & Tricritical mean field & \cite{LeGuillou} & Theory & 0.25 & 1.0 & 5\\
  MnSi && \cite{LeiZhang} & Modified Arrott plot & 0.242(6) & 0.915(3) & 4.734(6)
\end{tabular}
\end{ruledtabular}
\end{table*}

The critical exponents can also be determined by the Kouvel-Fisher (KF) method
\begin{equation}
\frac{M_s(T)}{dM_s(T)/dT} = \frac{T-T_c}{\beta},
\end{equation}
\begin{equation}
\frac{\chi_0^{-1}(T)}{d\chi_0^{-1}(T)/dT} = \frac{T-T_c}{\gamma},
\end{equation}
where $M_s(T)/[dM_s(T)/dT]$ and $\chi_0^{-1}(T)/[d\chi_0^{-1}(T)/dT]$ are linear in temperature with slopes of $1/\beta$ and $1/\gamma$, respectively.\cite{Kouvel} The linear fits give $\beta = 0.260(4)$, with $T_c = 60.05(13)$ K, and $\gamma = 1.136(6)$, with $T_c = 60.43(4)$ K, respectively [Fig. 4(b)].

To determine an appropriate model, the modified Arrott plots should be a set of parallel lines in high field region with the same slope [$S(T) = dM^{1/\beta}/d(H/M)^{1/\gamma}$]. Normalized slope ($NS$) is defined as $NS = S(T)/S(T_c)$, which enables us to determine the most suitable model by comparing it with the ideal value of unity. Plot of $NS$ vs $T$ for different models is shown in Fig. 5(a). One can see that the $NS$ of 3D Heisenberg model almost equals to unity above $T_c$, in accordance with the nearly isotropic magnetic character at high temperatures [Fig. 2(a)], while that of tricritical mean-field model is the best below $T_c$, indicating that the critical behavior of bulk CrI$_3$ may not belong to a single universality class. Figure 5(b) shows the isothermal magnetization $M(H)$ at a critical temperature $T_c$ = 60 K, with the inset plotted on a log-log scale. According to Eq. (3), the $M(H)$ at $T_c$ should be a straight line in log-log scale with the slope of $1/\delta$. Such a fitting yields $\delta = 5.32(2)$. In addition, $\delta$ can also be calculated from the Widom scaling law
\begin{equation}
\delta = 1+\frac{\gamma}{\beta}.
\end{equation}
From $\beta$ and $\gamma$ obtained with modified Arrott plot and Kouvel-Fisher plot, $\delta$ = 5.04(1) and $\delta$ = 5.37(4) are obtained, respectively, which are very close to that obtained from critical isotherm analysis. Therefore, the critical exponents $\beta$, $\gamma$, $\delta$, and $T_c$ obtained in the present study are self-consistent and accurately estimated within experimental precision.

Scaling analysis can be used to estimate the reliability of the obtained critical exponents and $T_c$. From Eq. (5), scaled $m$ vs scaled $h$ in linear and log-log scale [Fig. 6(a) and inset], all the data collapse on two separate branches below and above $T_c$. This can be also verified from plots of $m^2$ vs $h/m$, where it is seen that all data collapse on two different branches [Fig. 6(b)], confirming proper treatment of the critical regime. The scaling equation of state takes another form
\begin{equation}
\frac{H}{M^\delta} = k(\frac{\varepsilon}{H^{1/\beta}}),
\end{equation}
where $k(x)$ is the scaling function. From Eq. (10), all the experimental data should also fall into a single curve. This is indeed seen [inset in Fig. 6(b)]; the $MH^{-1/\delta}$ vs $\varepsilon H^{-1/(\beta\delta)}$ experimental data for CrI$_3$ collapse into a single curve and the $T_c$ locates at the zero point of the horizontal axis.

The obtained critical exponents of CrI$_3$, as well as the experimental values of Cr$_2$(Si/Ge)$_2$Te$_6$ and those of different theoretical models,\cite{LeGuillou, BJLiu, YuLiu, Zivkovic, LeiZhang, Kaul} are listed in Table I for comparison. Taroni \emph{et al.} have accomplished a comprehensive study of critical exponents for 2D magnets with a conclusion that the critical exponent $\beta$ for a 2D magnet should be within a window $\sim$ $0.1 \leq \beta \leq 0.25$.\cite{Taroni} In contrast to those of Cr$_2$(Si/Ge)$_2$Te$_6$ showing 2D Ising behavior coupled with a long-range interaction,\cite{BJLiu, YuLiu} the critical exponents of bulk CrI$_3$ crystals exhibit 3D critical phenomenon clearly. One can see that the critical exponents $\beta$ and $\gamma$ of CrI$_3$ lie between the theoretical values of 3D Ising model and tricritical mean-field model, suggesting that the interlayer coupling should not be neglected in bulk CrI$_3$. It is also interesting to mention that the $T_c$ of CrI$_3$ is the highest one in the CrX$_3$ (X = Cl, Br, and I) family.\cite{McGuire0, McGuire} Since the Cr-Cr distances increase with increasing halogen size, the direct exchange should weaken from Cl to Br to I. Therefore, the superexchange via Cr-X-Cr is expected to be FM and plays more important magnetic interaction.\cite{Lado} Moving form Cl to Br to I, more and more covalent Cr-X bonds strengthen superexchange interactions and raise ordering temperatures, as well as increase spin-orbital coupling, which may account for its large magnetic anisotropy.\cite{Lado}

Then it is important to understand the nature as well as the range of interaction in this material. In a homogeneous magnet the universality class of the magnetic phase transition depends on the exchange distance $J(r)$. In renormalization group theory analysis the interaction decays with distance $r$ as
\begin{equation}
J(r) \approx r^{-(3+\sigma)},
\end{equation}
where $\sigma$ is a positive constant.\cite{Fisher1972} Moreover, the susceptibility exponent $\gamma$ is predicted as
\begin{multline}
\gamma = 1+\frac{4}{d}(\frac{n+2}{n+8})\Delta\sigma+\frac{8(n+2)(n-4)}{d^2(n+8)^2}\\\times[1+\frac{2G(\frac{d}{2})(7n+20)}{(n-4)(n+8)}]\Delta\sigma^2,
\end{multline}
where $\Delta\sigma = (\sigma-\frac{d}{2})$ and $G(\frac{d}{2})=3-\frac{1}{4}(\frac{d}{2})^2$, $n$ is the spin dimensionality.\cite{Fischer} When $\sigma > 2$, the Heisenberg model is valid for 3D isotropic magnet, where $J(r)$ decreases faster than $r^{-5}$. When $\sigma \leq 3/2$, the mean-field model is satisfied, expecting that $J(r)$ decreases slower than $r^{-4.5}$. In the present case, it is found that the magnetic exchange distance decays as  $J(r)\approx r^{-4.69}$, which lies between that of 3D Heisenberg model and mean-field model.

\section{CONCLUSIONS}

In summary, we have made a comprehensive study of the critical behavior around the PM-FM transition in bulk CrI$_3$. The PM-FM transition in CrI$_3$ is identified to be the second order in nature. The critical exponents $\beta$, $\gamma$, and $\delta$ estimated from various techniques match reasonably well and follow the scaling equation, suggesting a 3D long-range magnetic coupling with the exchange distance decaying as $J(r)\approx r^{-4.69}$.

\section*{Acknowledgements}
We thank John Warren for help with the scanning electron microscopy (SEM) measurements. This work was supported by the US DOE-BES, Division of Materials Science and Engineering, under Contract No. DE-SC0012704 (BNL).

\end{document}